\documentclass{desyproc}

\usepackage{textcomp}

\begin{document}
%------------------------------------
\title{ALPS\,II technical overview and status report}

%for single authors the superscripts are optional
\author{{\slshape  Aaron Spector$^1$ for the ALPS collaboration$^{\dagger}$ }\\[1ex]
$^1$Institut f\"ur Experimentalphysik, Universit\"at Hamburg, Hamburg, Germany\\
$^\dagger$https://alps.desy.de/e154/}

% if the proceedings are available online (e.g. at Indico)
% please enter the contribution ID or file_name below for the DOI
%\contribID{32}
\contribID{spector\_aaron}

% TO THE CONFERENCE EDITORS: 
% please update the following information      
% before sending the template to the authors
\confID{13889}  % if the conference is on Indico uncomment this line
\desyproc{DESY-PROC-2016-XX}
\acronym{Patras 2016} % if you want the Acronym in the page footer uncomment this line
\doi  % if there is an online version we will register DOIs

\maketitle

\begin{abstract}

The Any Light Particle Search II (ALPS\,II) is an experiment that utilizes the concept of resonant enhancement to improve on the sensitivity of traditional light shining through a wall style experiments. 
These experiments attempt to detect photons passing through an opaque wall by converting to relativistic weakly interacting sub-eV particles and then reconverting back to photons.
ALPS\,II at DESY in Hamburg, Germany will use dually resonant optical cavities before and after the wall to increase the probability of this interaction occurring. This paper gives a technical overview and status report of the experiment.

\end{abstract}

\section{Introduction}

The Any Light Particle Search II (ALPS\,II) is a light shining through a wall experiment that will search for WISPs (Weakly Interacting Sub-eV Particles) in the mass range below 1\,meV~\cite{alpstdr}. Experiments of this type feature a light source shining at an opaque wall or optical barrier and then attempt to observe this light passing through the barrier. A few of the photons from the light source will convert to relativistic WISPs which then traverse the wall and after the wall some of these WISPs will then reconvert back to photons to create a measurable signal. For some classes of particles a magnetic field is necessary for these interactions to take place.

A previous light shining through a wall experiment, ALPS\,I, featured a Production Cavity (PC) to increase the number of photons circulating in the region before the wall. ALPS\,II will be the first experiment to use a Regeneration Cavity (RC) to resonantly enhance the probability that WISPs will reconvert back to photons after the wall~\cite{rc}. This requires that the PC and RC share the same resonant frequency and Eigenmode such that 95\% of the light from the PC would couple to the RC if it were directly incident on it. This must be accomplished without allowing any of the photons circulating inside the PC to breach the wall and enter the regeneration side of the experiment as they will be indistinguishable from the signal photons we are attempting to measure. This paper gives a brief overview and technical status report for ALPS\,II.

%\subsection{Sensitivity to WISPs}

\section{Experimental design}

ALPS\,II is currently being constructed in two stages. The first stage, ALPS\,IIa, will feature 10\,m cavities with no magnets to test the optical subsystems related to maintaining the dual resonance of the PC and RC as well as single photon detection schemes at the output of the RC. Even though ALPS\,IIa will not be sensitive to axion-like particles due to the lack of magnets, it will represent the most sensitive search to date for hidden photons in the mass range $m_{\gamma'}>5\times10^{-5}$\,eV~\cite{alpstdr}. ALPS\,IIc, the second stage, represents the full scale experiment using 100\,m cavities with light propagating through high magnetic fields.

\begin{figure}[]
\centerline{\includegraphics[]{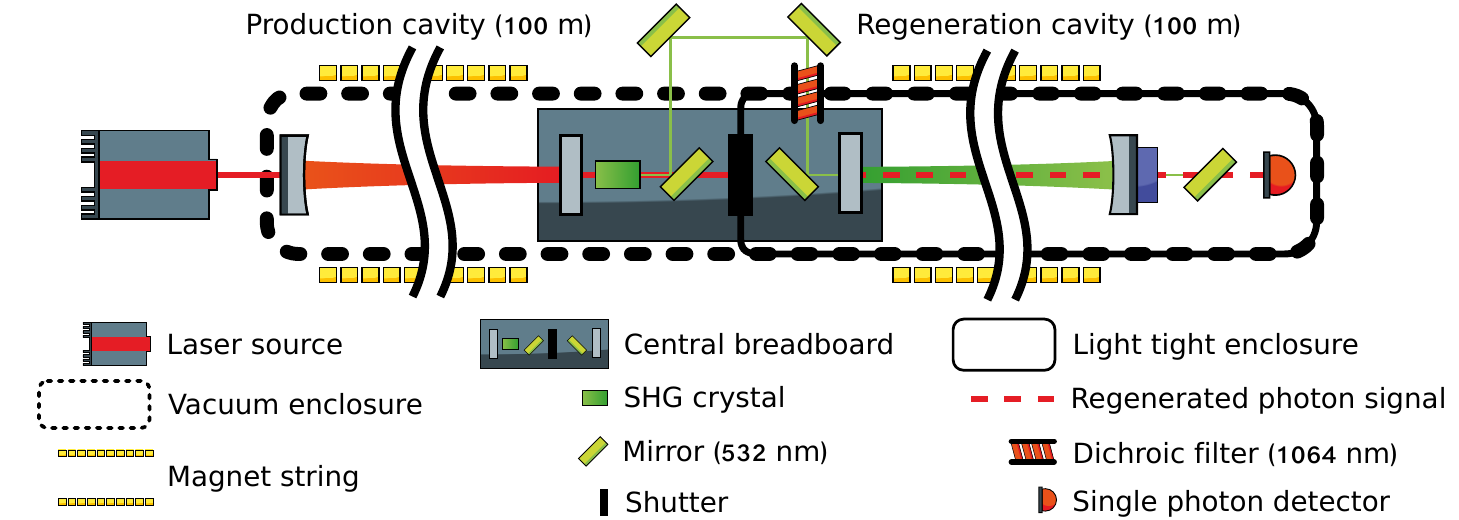}}
\caption{ALPS\,IIc simplified design.}\label{Fig:2c}
\label{sec:figures}
\end{figure}

A simplified diagram of the ALPS\,IIc design is shown in Figure~\ref{Fig:2c}. The PC can be injected with up to 35\,W of 1064\,nm light from a laser system composed of a non-planar ring oscillator with a Nd:YVO$_4$ amplification stage. With a power buildup factor of roughly 5000 there will be 150\,kW of power circulating in the PC.  A feedback control system will maintain the resonance condition of laser with respect to PC length by actuating on the laser frequency.

The RC will have a power buildup factor of 40,000 for 1064\,nm light. Since none of the light circulating in the PC is allowed to couple into the RC a Second Harmonic Generation (SHG) crystal will be used to frequency double the light at the output of the PC. The 1064\,nm light is then removed from this beam with a sophisticated system of dichroic filters. This light is coupled into the RC and used in a feedback control scheme that will stabilize its length via a piezo electric device mounted to the out-coupling mirror.

%It is essential for the RC to maintain the same resonant frequency as the PC to enhance the reconversion probability of the WISPs generated by the PC. %While maintaining the dual resonance of the two cavities the length stabilization control loop for the 
The resonant frequency of the RC  must be within 2\,Hz of the frequency of the light circulating inside the PC to ensure the resonant enhancement of the reconversion probability of WISPs generated by the PC. This is equivalent to the position of this mirror being actively controlled to better than a picometer.

The PC and RC must also occupy the same spatial Eigenmode. For ALPS\,IIc both cavities will be operate in a nearly semi-confocal configuration with end mirrors designed to have a radius of curvature of 250\,m and flat mirrors located at the waist of the Eigenmode. As shown in Figure~\ref{Fig:2c} both of the flat mirrors will be mounted directly to the Central Breadboard (CBB) to ensure that there are no alignment drifts between these mirrors. ALPS\,IIa also operates both cavities in a nearly semi-confocal configuration.% however because of the shorter length the end mirrors have a radius of curvature of 23.5\,m.

The light tight enclosure for ALPS\,IIc and ALPS\,IIa will be equipped with a shutter along the optical axis of the cavities. %When the shutter is open light transmitted through the PC will be directly incident on the RC. 
This will allow us to open the shutter and check the spatial overlap between the PC and RC as well as the dual resonance.  %We must be able to couple 95\% of this light to the RC in order for ALPS\,IIc to meet its requirements on the resonant enhancement of the reconversion probability.

ALPS\,II will need to measure the regenerated photon signal corresponding to single photons over the course of several days. This will be done with two independent techniques. One uses a transition edge sensor (TES) that consists of a superconducting tungsten chip, a constant voltage source and an inductive coil \cite{Bast}. The chip is linked to a thermal bath that maintains its temperature at the superconducting transition. Signal photons from the RC are directed to the chip through fiber optics and when a photon is incident on the chip it is absorbed and causes a sudden rise in temperature. This temperature rise in the chip causes a rise in its resistance and thus a drop in the current through the circuit. This negative pulse in the current is read out with a double stage superconducting quantum interference device. 

The other method is a heterodyne detection scheme that optically interferes the measurement signal with a local oscillator and then measures the interference beatnote with a photodetector. The phase relation between the local oscillator and the measurement signal will be tracked at a separate point in the experiment and then this information will be used to sum coherently the measurement signal. 

ALPS\,IIc will use 5.3\,T superconducting dipole magnets from the decommissioned HERA accelerator to generate the magnetic field. With a string of ten magnets for each cavity, their respective Eigenmodes will each propagate through 466\,T$\cdot$m of magnetic field length. 

Equation~\ref{EQ} gives an expression for the sensitivity of ALPS\,IIc to the coupling constant between axion-like-particles and photons, $S(g_{\alpha\gamma})$. Here $BL$ is the magnetic field length while $t$ is the measurement time. The detector background rate is given by $D_n$ while its detection efficiency is $\eta$. $P_i$ is the power incident on the PC while $\mathcal{F}_{PC}$ and $\mathcal{F}_{RC}$ represent the power buildup factors of the PC and RC.
\begin{equation}\label{EQ}
S\left(g_{\alpha\gamma}\right)\propto \frac{1}{BL}\left(\frac{D_n}{t}\right)^{\frac{1}{8}}\left(\frac{1}{\eta\, P_{i}\,\mathcal{F}_{PC}\mathcal{F}_{RC}}\right)^{\frac{1}{4}}
\end{equation}
With these systems ALPS\,IIc will have the sensitivity either to detect or rule out axion-like particles with a coupling constant of $g_{a\gamma}\gtrsim2\times10^{-11}$\,GeV$^{-1}$ for masses below $10^{-4}$\,eV~\cite{alpstdr}.

\section{Status}

A 20\,m cavity was built to characterize several of the input optics control systems for ALPS\,IIa. The cavity used two of the in-coupling mirrors for the PC in a nearly confocal configuration that would produce the same Eigenmode as the ALPS\,IIa PC and RC. This allowed us to construct and test the frequency stabilization, automatic alignment control system, and power stabilization scheme. Together these systems were able to reduce the RMS of relative power noise in transmission of the cavity to better than $4.0\times10^{-5}$ \cite{20m}.

The frequency stabilization scheme suppressed the frequency noise between the input laser and the cavity resonance up to a unity gain frequency of 55\,kHz. The control system proved to be very robust and was able to maintain the cavity resonance for up to 48\,hours before being manually stopped. 
An automatic alignment system monitored and controlled the relative pointing between the cavity Eigenmode and the input beam using quadrant photodetectors to do differential wavefront sensing. This ensured that the maximum possible power from the input laser couples to the cavity. 
% and was able to suppress alignment fluctuations below 15 Hz.
 In the absence of intra-cavity losses the 20\,m cavity was expected to have a power buildup factor of 1,140. We were able to achieve a power build-up factor of 1,010 demonstrating that our systems are capable of maintaining the resonance condition for a long-baseline low-loss optical cavity.

%A test PC was assembled using a flat mirror rigidly mount to a CBB prototype. We demonstrated that it was possible to align the test PC by using the current alignment system for the CBB. The laser was locked and we were able to achieve a power buildup of roughly 3000 with 500\,mW incident on the cavity. The laser power incident on the cavity was then increased to 30\,W and we achieved 50\,kW of circulating power in the cavity.

The CBB is currently being built at the Albert Einstein Institute in Hannover, Germany. Tests there have demonstrated that it is possible to align the PC and RC mirrors on the CBB with better than 5\,{\textmu}rad accuracy. Furthermore, long term measurements suggest that alignment drifts meet our requirements as the angular misalignment between the two mirrors never exceeded the 10\,{\textmu}rad ALPS\,IIa requirements.

%20 m Cavity was built. Suppress power fluctuation in transmission.

%An automatic alignment system was built.

%PC built with mirror on CBB prototype, were able to align cavity.

%50 kW of circulating power has been achieved in the PC.

%CBB is being manufactured in Hannover

%\section{Detection Schemes}

%Two separate detection schemes are currently being investigated by the ALPS collaborations. A TES is currently being characterized at DESY. The University of Florida is developing a heterodyne detection scheme which uses a second laser as a local oscillator to perform an AC measurement of the detection signal.

%\subsection{Transition Edge Sensor}

%The TES consist of an absorptive tungsten chip in series with a constant voltage source and an inductive coils. The chip is linked to a thermal bath that maintains its temperature at the critical temperature at which it transitions to being superconducting. Then when a photon hits the chip it is absorbed and causes the chip to suddenly rise in temperature. This temperature rise causes a rise in the resistance of the chip and thus a drop in the current flowing through the inductive coil. This negative pulse in the current through the inductive coil is read out with a SQUID. 

An evaluation of the detection efficiency and energy resolution of the TES is underway and preliminary results suggest a detection efficiency better than 30\% and a energy resolution better than 0.1\,eV~\cite{Bast}.
A similar system at NIST showed a detection efficiency better than 95\% as well as an energy resolution of  0.29\,eV for photons at a wavelength of 1556\,nm~\cite{TES}. 
As Equation~\ref{EQ} shows limiting the background event rate is essential to optimizing the sensitivity of ALPS\,IIc. The primary source of background is expected to be photons generated by blackbody radiation from the optics and fiber system which direct the measurement signal to the TES. %While the rate of  photons with an energy of 1.16\,eV (1064\,nm) due to blackbody radiation is expected to be very low, it is very difficult to distinguish the signal generated by pileups of two coincident photons whose energies sum to 1.16\,eV and the signal from the 1064\,nm photons that we are attempting to measure. 
An algorithm which fits a theoretical pulse shape to triggered events to help identify signal pulses from pileups has been implemented along with a faster readout system which is expected to improve its precision. % Its effect on the dark count rate is currently being assessed. 
 Additionally, we plan to install a fiber bandpass filter situated in a cold environment to reduce the rate of blackbody photons incident on the TES.
For the heterodyne detection scheme a test-bed is being set up at the University of Florida. 
Results from this setup are expected in the coming months.
 %There a digital system is under construction to track the phase of the measurement signal with respect to the local oscillator and then use this information to coherently sum the measurement signal. 
%Simulations have shown that the system is capable of integration times of up to 1000\,s with no appearance of spurious signals.

%Since the dipole magnets for ALPS\,IIc were originally to steer particle beams around the curves of the HERA ring, each has a 600\,m radius of curvature. 
The magnets for ALPS\,IIc each have a radius of curvature of 600\,m and must be unbent before being operated to create a large enough aperture for the light circulating inside of the optical cavities. The full experiment requires 20 magnets. So far two have been unbent and both were successfully operated.

\section{ALPS\,II Timeline}

The PC for ALPS\,IIa is now in the final stages of commissioning and we will begin installing the CBB and input optics for the RC by the end of 2016. We expect to perform a hidden photon search with ALPS\,IIa within the next year. Work will begin preparing the HERA tunnel and constructing the clean rooms for ALPS\,IIc by the end of 2017 and the unbending of the magnets is scheduled to be completed in  2018. This will allow for the commissioning of the optical systems for ALPS\,IIc to take place in 2019 with measurement runs beginning in 2020.

\section*{Acknowledgments}

The authors would like to thank the DFG and particularly the SFB 676, as well as the Heising-Simons Foundation for funding support.

% ****************************************************************************
% BIBLIOGRAPHY AREA
% ****************************************************************************

\begin{footnotesize}

\end{footnotesize}

% ****************************************************************************
% END OF BIBLIOGRAPHY AREA
% ****************************************************************************

\end{document}